# Dilemma of the Artificial Intelligence Regulatory Landscape

Weiyue Wu and Shaoshan Liu
PerceptIn

**Abstract:** As an autonomous driving start-up company, we have struggled with a broad spectrum of regulatory requirements for over four years. A large amount of time, effort, and budget have been poured into varying compliance procedures instead of R&D. In a way, the complex and varying regulatory processes subtly give an advantage to well-established and resourceful technology firms over resource-constrained AI start-ups. However, our situation is not alone and, to some extent, reflects the dilemma of the artificial intelligence (AI) regulatory landscape. This article introduces our first-hand experiences dealing with the varying AI regulatory frameworks and provides practical recommendations for AI start-ups to work with regulators efficiently and effectively.

## 1. Introduction

When legal regulations get ahead of technological developments, the progress of a human society may be constrained. Conversely, when technological developments run ahead of legal regulations, unregulated new technologies may harm human society, defying technological development's fundamental purpose.

This is exactly what has happened in our world in the past decade, as technological developments far outpaced legal regulations. Worse, traditional legal frameworks focus on the relationship between people, whereas we must develop a legal framework to regulate relations between people and intelligent machines in the current era. Integrating AI technologies into human society imposes unique legal challenges without any precedence.

For the first time in history, AI has the potential to generate solutions that are superior to what a human would expect [1]. However, today AI is not bound by ethics, and decisions made by AI may be considered out of line with standards and ethics generally accepted by human society [2]. The unforeseeable superiority and the problem of ethical control may create chaos in the AI regulatory landscape. If AI is not adequately regulated, such problems will further exacerbate unforeseen social issues that the current legal framework is unprepared for.

As an autonomous driving start-up founded in 2016, we have launched commercial autonomous driving operations in the U.S., Europe, Japan, and China to find the most promising market [3]. It has been a treacherous journey filled with regulatory roadblocks and land mines. Such a problem is not unique to autonomous driving but universal to all AI applications. In this article, we share our first-hand experiences tackling regulatory requirements and attempting to reduce the friction between innovative technologies and regulatory requirements to advance human society with properly regulated AI technologies.

## 2. Our Global Commercial Deployment Experiences

As an autonomous driving start-up, our business model is to provide turn-key solutions to customers. The solution includes hardware such as drive-by-wire chassis and computing units, a full stack of autonomous driving software such as perception and localization, and a user interface [4].

In the past four years, when we performed commercial deployments in different regions worldwide, we bumped into a broad spectrum of regulatory rules across countries or regions within a country. This lack of regulation standardization has become unhealthy as companies in the AI sector, mostly start-ups, have to spend a significant amount of budget, time, and effort dealing with different regulatory measures. This section summarizes our first-hand experiences in working with authorities worldwide.

## 2.1 Deployments in China

In 2017, autonomous driving test permits in China were virtually non-existent. Only a few governmental documents stated high-level goals and principles in some pilot cities like Shanghai. The absence of test permit certification pushed start-ups in this field to test their vehicles wildly on highways, where the traffic was more predictable and less complicated, or in rural areas where the population is scarce, both posing fatal risks to the general public.

To minimize compliance and safety risks, we chose to carry out autonomous driving projects in restricted areas, such as factories and school campuses. Since these areas are private, the test pod could go on the roads once approved by the property management. In addition, we set the speed limit at 20 miles/hour to cope with safety risks and designed several redundant perception systems to avoid accidents [5].

Having accumulated 1.5 months of testing data and different driving scenarios, our company was invited to join an autonomous driving pilot program in Shenzhen, China's Silicon Valley. We were selected to present the latest autonomous driving technology and global trend in this field in a seminar hosted by the local government, where we seized the opportunity to emphasize the urgency and importance of regulated test drives. After six months, our voice was heard, and we could file our first test permit application. In this case, the whole regulatory process lagged behind the actual technological development, pushing companies to play wild first and file applications later. Furthermore, if a company goes to another city for a new autonomous driving project, the whole process will start from zero.

## 2.2 Deployments in Europe Union

In late 2018 we were invited to carry out autonomous driving operations in a European city. It took us over seven months to complete the application process and launch commercial deployment. The first three months focused on technical specifications such as communication bandwidth requirements and visual perception response time. The latter four months focused on functional specifications in more than 40 scenarios, from recognizing an unexpected object in the test field to preparing a backup test plan on rainy days in the dry city. Filing all requested reports, we were granted a restrictive operation permit that lasted a month. In Europe we were strictly not allowed to perform any experiments on the roads, whether public or private, without a valid pemit. In addition, there was no room for negotiation between local authorities and us.

Regardless of the actual operation environment, we had to spend extended amount of time and effort to prepare for all scenarios and go through the regulatory process. This imposes a significant financial burden on AI start-ups, which are already constrained on budget and race against time for growth. In a way, the complex and varying regulatory processes across different regions subtly give advantage to well-established and resourceful technology firm over resource-constrained start-ups.

## 2.3 Deployments in Japan

With autonomous driving cases launched in China and Europe, we entered Japan market and directly worked with Japan's Ministry of Land, Infrastructure, Transport, and Tourism (MLIT) to carry out a driverless vehicle project [6]. The experience in Japan was a combination of China and Europe. We could negotiate with authorities to start from the most straightforward scenarios and reduce paperwork. Meanwhile, the authority closely reviewed technical specifications and risk-mitigation plans in those scenarios.

The application and endorsement process is top-down. We had to work with the national government to gain approval and then trickled down to local governments to carry out the operation. We first ran the test operation in a park in Fukuoka under the supervision of MLIT. The operation data were then used to go through a basic safety check to assess the risk of our autonomous driving operations. After the procedure was completed, we were allowed to continue our operation at the new site in a historic park in Nara. The whole operation scale-up process took more than 12 months, and it was worthwhile since we did not have to repeat the initial certification process once we received endorsements from MLIT.

## 2.4 Deployments in the U.S.

In contrast to Japan, we had to take a bottom-up approach for deployments in the U.S. The local governments are responsible for inner-city transportation, and these local governments have the authority to grant autonomous driving operation permits. Since the state and the federal governments had limited guidance on the deployment of autonomous driving services, we had to deal with each city individually to expand our operation, which is an inefficient and costly process given the regulatory differences among cities. In our cases in the U.S., the test data and safety assurance plans in California did not give us privileges when we applied for the test permit in Indiana. Although such an arrangement may be adequate to mitigate people's safety risk in different states with different traffic densities [7], there could be room to improve process standardization and transparency, which would save AI start-ups' resources and, in turn, help advance the technology.

## 3. The Dilemma

Figure 1 summarizes the basic process of autonomous driving deployment in the four areas discussed above. The deployment process consisted of four essential parts: Application filing to initiate the conversations, national-level approval, local-level approval, and commercial deployment. There was no explicit instruction to follow, and it took at least eight months to reach commercial deployment after technical readiness. In addition, the broad spectrum and the opacity of regulatory requirements consumed a significant amount of the company's resources, costing 42% of the annual budget to deal with regulatory issues, compared with the average 13% of the budget in software companies [8].

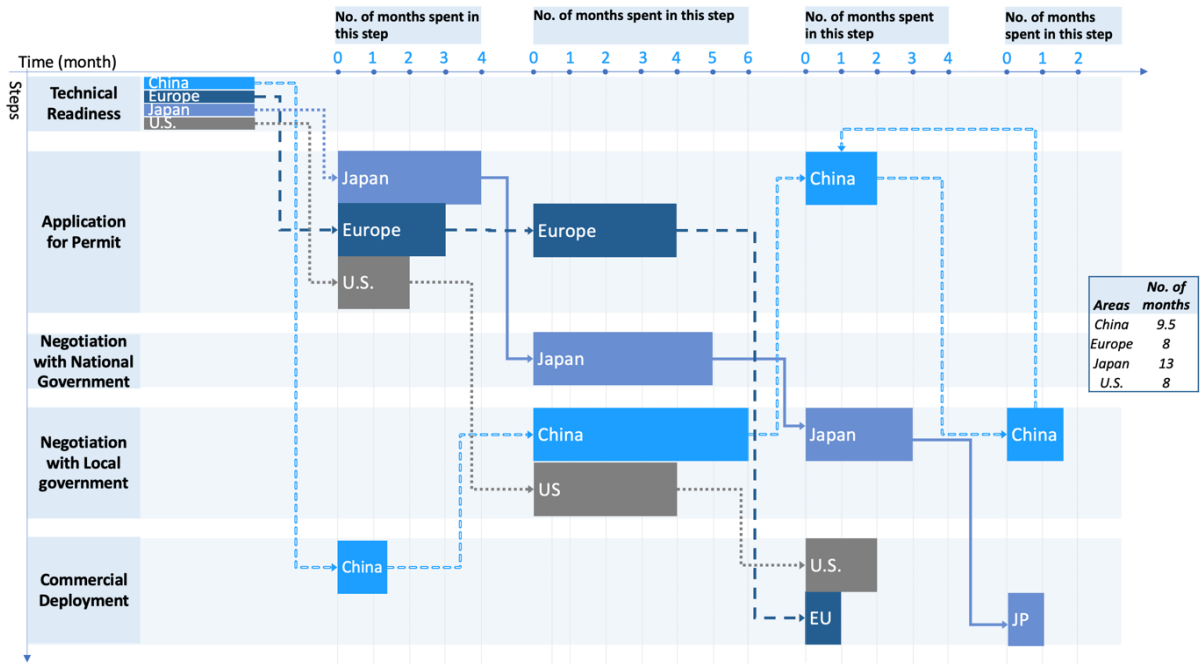

Figure 1: Process of getting a test permit for autonomous driving in different regions

Traditionally software companies may take an incremental approach to comply with regulations. They first launch a minimum viable product, collect feedback from pilot users, and then reiterate several times until the product rolls out to the general public. Then more compliance issues may arise yet get addressed on the fly with software updates.

Conversely, autonomous driving start-ups have to deal with various compliance issues from day one, which imposes a significant financial burden on autonomous driving start-ups. Dealing with regulatory uncertainty is a ubiquitous problem among many AI companies, as exemplified by Microsoft's retracting its facial recognition products [9]. We believe this dilemma is caused by a lack of standardization in the AI product deployment process and a lack of understanding of AI technologies in the regulatory bodies.

Without a properly regulated environment, AI companies cannot thrive and transform their latest technology into successful products. When regulatory requirements are absent, players in the field may go wild and put both the company's future and public safety at stake, jeopardizing technological advancement in the long run. When the requirements are too tight, AI companies, especially those start-ups which do not have sufficient budget for compliance, would be exhausted with filing repetitive paperwork to different offices.

## 4. Recommendations for AI Start-ups

Our lessons can help AI companies, especially start-ups with limited resources, to navigate through various regulatory barriers. While the specific questions being asked by regulators varied from time to time, we have found that the key to settling concerns is to clearly convey the message that prospectus benefits outpasses relevant risks. These recommendations have been proven effective in our own deployment projects across multiple countries.

First, cross-function communication and information sharing are critical to unifying the team to deal with external complexities, as complex regulatory requirements often cause chaos within an organization. For instance, as we carried out deployment projects, our field engineers often needed to work directly with regulators to demonstrate technical progress, fix deployment problems, and address their concerns. Engineers with cross-functional experiences tend to outperform those without, as the former ones are capable of communicating effectively with regulators who have limited technical background.

Second, show strong evidence to support the benefits of the project. Most regulators welcome innovations, and they are eager to integrate innovative technologies into their economies. They look for strong evidence that the technology indeed works, and they look for evidence from credible sources, such as publications from leading journals in the field or reports released by leading research institutes. Hence, we suggest AI start-up companies collaborate with academic partners to perform scientific research to verify the project's benefits. Industry-academia collaboration is essential to foster the AI industry more than ever before.

Third, present a concrete risk mitigation plan to regulators. Most regulatory actions are designed to mitigate risks. The local authority will often carry the blame if an innovative project goes wrong. Therefore, regulators would try their best to mitigate risks. Presenting a clear risk mitigation plan and success stories of deployment case studies elsewhere will undoubtedly improve the regulator's confidence in the project. While in our own experiences, presenting the cases in China and Europe helped us obtain approval for operation in Japan. For instance, development and awareness of technical standards, such as Safety Of The Intended Functionality (SOTIF) For Autonomous Driving [10], is an effective method to improve product creditability and to boost regulators' confidence.

## 5. Summary

To summarize, our first-hand experiences of dealing with a broad spectrum of regulatory requirements pinpoint the dilemma of the AI regulatory landscape today. We provide some practical recommendations for AI start-ups to work efficiently and effectively within the regulatory framework to address this problem. However, these recommendations only make the AI start-ups better adapt to the regulatory landscape. We are hoping for standardization in the AI regulatory process, which will make the AI industry much more efficient and effective.